\documentclass[12pt]{iopart}

\usepackage{graphicx}  

\begin{document}

\title[Adiabatic coupling in waveguide structures]{Using adiabatic
  coupling techniques in atom-chip waveguide structures}

\author{B.~O'Sullivan, P.~Morrissey, T.~Morgan and Th.~Busch}

\address{Physics Department, University College Cork, Cork, Ireland}
\ead{bosullivan@phys.ucc.ie}

\begin{abstract} Adiabatic techniques are well known tools in
  multi-level electron systems to transfer population between
  different states with high fidelity. Recently it has been realised
  that these ideas can also be used in ultra-cold atom systems to
  achieve coherent manipulation of the atomic centre-of-mass
  states. Here we present an investigation into a realistic setup
  using three atomic waveguides created on top of an atom chip and
  show that such systems hold large potential for the observation of
  adiabatic phenomena in experiments.
\end{abstract}

%\pacs{00.00, 20.00, 42.10}

\section{Introduction}
\label{sec:Introduction}

Trapping and controlling small numbers of neutral atoms has, in recent
years, emerged as one of the most active and productive areas in
physics research
\cite{Haensel:01,Bergamini:04,Chuu:05,Shevchenko:06}. Such systems
allow to perform experiments to answer fundamental questions in
quantum mechanics \cite{Eschner:01,Beugnon:06} and hold great
potential for use in quantum information processing
\cite{Brennen:99,Jaksch:99,Jaksch:00}.  Advances in the technology of
optical lattices and micro-traps have allowed for substantial progress
in this area
\cite{Birkl:01,Rauschenbeutel:05,Yavuz:06,Sortais:06,Fortier:07} and
various concepts have been developed to prepare and process the states
of single atoms. While techniques for controlling and preparing the
internal states of atoms using appropriate electromagnetic fields are
well developed, only a few concepts exist for achieving the same
control over the spatial part of a wavefunction
\cite{Zhang:06,Eckert:04, Eckert:06, Jaksch:99,Karski:09}. Such
control would complement currently existing techniques and allow for
the complete engineering of a particle's quantum state.

One area where control over the spatial part of the wavefunction is
important is the challenge of devising techniques for controlled
movement of atoms between different regions in space. In optical
lattices this corresponds to moving between discrete lattice sites and
in waveguide settings this would allow transfer from one guide to
another. Direct tunneling is a coherent process that can achieve this,
however Rabi-type oscillations make it experimentally very hard to
reach high fidelities \cite{Mompart:03}.

Recently it was pointed out that systems consisting of three separated
centre-of-mass modes allow for the use of STIRAP-like processes to
achieve robust transfer of atoms from one position to another with
high fidelity \cite{Eckert:04, Eckert:06, Greentree:04}. The process
of Stimulated Raman Adiabatic Passage (STIRAP) is well known in
three-level-optics, where it refers to the technique of applying a
counter-intuitive sequence of laser pulses to achieve a transition of
an electron between the two ground states in a $\Lambda$-system
\cite{Bergmann:98,Vitanov:01}. In the atom trap scenario the energy
levels are replaced by spatially separated trap ground states and the
laser interaction is replaced by the coherent tunneling interaction.

One advantage of adiabatic techniques is their large robustness
against experimental uncertainties as long as the whole process is
carried out mostly adiabatically \cite{Harkonen:06}. However, this
also means that a resonance between the asymptotic eigenstates has to
exist, which is a condition that for many realistic situations is hard
to ensure. Suggestions for and examinations of realistic systems in
which the STIRAP process could be observed for cold atomic gases are
therefore currently very rare.

In this work we will focus on atom-chip systems and investigate their
suitability to observe this adiabatic process.  These micro-fabricated
chips, on which surface mounted, current carrying wires provide
guiding potentials for matter waves, can be loaded with ultracold atom
gases at low densities. As opposed to traditional experimental setups,
these systems allow reaching smaller dimensions and the wire geometry,
and therefore the waveguide geometry, can be chosen almost at
will \cite{Folman:02}.

The first investigation into adiabatic techniques in waveguides was
presented by Eckert {\it et al.}  \cite{Eckert:06}, who showed that a
CPT-like process which acts like a 50:50 beam splitter could be
realised with a large degree of fidelity. While the initial state for
a numerical evolution can be prepared with a large degree of
localisation, one of the problems following the subsequent evolution
inside the waveguide is that the wavefunction disperses along the
guide. This makes it hard to exactly measure the final state of the
system and put a quantitative number on the efficiency of the
adiabatic process. Here we will introduce a simple harmonic potential
along the longitudinal direction of the trap, which will allow us to
perfectly measure the fidelity of the process. It is also worth
mentioning that STIRAP in optical waveguides with classical light has
been observed recently \cite{Longhi:07}.

In the next section we will first remind the reader by briefly
reviewing the idea of STIRAP and its translation into the realm of
waveguides. After that, in Section \ref{sec:Model}, we will examine a
model waveguide potential in which the resonance condition is
fulfilled throughout the whole process and show that the dispersion of
the wavefunction in the longitudinal direction has no significant
effect on the fidelity of the process. In Section \ref{sec:AtomChips}
we describe a realistic situation by examining three waveguides
created on top of an atom chip. We show that even though the resonance
condition is not fulfilled at all times, a counter-intuitive approach
will lead to larger transfer and can clearly be distinguished from a
direct tunneling approach. Finally we conclude.

\section{STIRAP}
\label{sec:STIRAP}

In this section we will briefly review the basic idea of STIRAP, which
is a technique originally developed for transitions in optical
$\lambda$-systems and which makes use of a two-photon Raman
process. By applying the pump and the Stokes pulse in a
counter-intuitive time-ordered way it leads to population transfer
directly from one of the ground states to the other without any
population ever being in the excited transitional state. In optical
systems this inhibits spontaneous emission and is therefore
often referred to as a dark-state technique.

The basic idea can be understood in the simple model of a three state
system described by the Hamiltonian
\begin{equation}
  \label{eq:Hamiltonian}
  H(t)=\hbar
    \left(\begin{array}{cccc}
    0          & -\Omega_P(t) & 0  \\
    -\Omega_P(t)  & 0 & -\Omega_S(t) \\
    0 & -\Omega_S(t) & 0  
\end{array}\right) \;,
\end{equation}
where we have set the energies of the three asymptotic eigenstates to
zero and the Rabi frequencies of the pump and the Stokes pulses are
given by $\Omega_P$ and $\Omega_S$, respectively. This Hamiltonian can
be diagonalised and the eigenstate which is of interest to us here is
the so-called dark state given by
\begin{equation}
  |d\rangle=\cos\theta|1\rangle-\sin\theta|3\rangle\;,
\end{equation}
where the mixing angle $\theta$ is given by
$\tan\theta=\Omega_P/\Omega_S$. This angle describes how the the
population is distributed between the two states $|1\rangle$ and
$|3\rangle$ and it can be chosen by varying the strength of the pump
and the Stokes pulse with respect to each other in time. In
particular, if the intensity of the Stokes pulse increases before that
of the pump pulse (counter-intuitive coupling scheme), one finds that
all initial population in $|1\rangle$ will be transferred to
$|3\rangle$.

The fact that this process can be observed for trapped atoms was first
pointed out by Eckert {\it et al.} \cite{Eckert:04}. The asymptotic
eigenstates of the Hamiltonian (\ref{eq:Hamiltonian}) are then the
spatial modes the atoms occupy and the time-dependent coupling is
given by the tunneling strength between these modes. While the
time-dependence of the tunneling strength can be achieved by
temporally changing the distance or the barrier height between the
individual states, an atom moving in a waveguide can also experience
this as a function of travelled distance \cite{Eckert:06}. In the next
section we will examine an example of this.

\section{Model}
\label{sec:Model}

The Schr\"odinger equation for the evolution of a wave-packet in a
two-dimensional waveguide structure is given by
\begin{equation}
  i\dot\psi(x,y) =
  -\frac{\hbar^2}{2m}\nabla^2\psi(x,y)+V(x,y)\psi(x,y)\;,
  \label{eq:SE}
\end{equation}
where $m$ is the mass of the atom. As the third dimension does not
significantly contribute to the dynamics we are aiming to observe, the
restriction of the above Hamiltonian to two dimensions is
justified. In this section we will first examine the STIRAP process
using an idealised potential in which the condition of resonance
between the individual waveguides is fulfilled at any point. This will
help us to illustrate the basic process and in particular highlight
the influence of the longitudinal dimension. In Section
\ref{sec:AtomChips} we will compare these results to realistic atom
chip scenarios in which we will have to relax the resonance condition.

\begin{figure}[t]
\centering
  \includegraphics[width=0.5\linewidth]{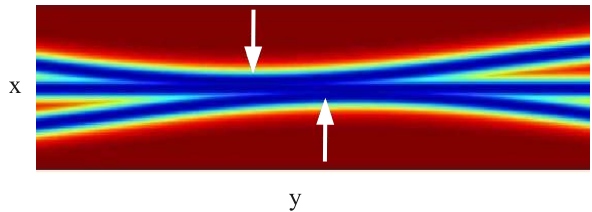}
  \caption{Waveguide structure near the point of closest approach. The
    points where the upper and lower waveguides have a minimum
    distance from the central waveguide are indicated by the
    arrows. The wave-packet will originally travel in the lowest
    waveguide from left to right.}
  \label{fig:Schematic}
\end{figure}

The assumption we make to guarantee that the ground state energy in
all three waveguides is the same everywhere is that we can construct
our potential $V(x,y)$ by stitching three independent waveguides
together. In a realistic situation the potentials creating each guide
would influence each other and lead to non-symmetric situations
between pairs. We assume each guide to have the potential
\begin{equation}
  V_s=A\tanh[B(x-f(y))]^2
\end{equation}
where $A$ determines the height, $B$ the width and $f(y)$ the position
of the minimum along the $x$-axis. The overall potential is then
assumed to be given by the minimum value of each of the three
potentials at any point in space. A schematic view of the area in
which the guides approach most closely is shown in
Fig.~\ref{fig:Schematic}.

The eigenstates of matter waves propagating in two-dimensional
waveguides at different distances have recently been explored by
J\"a\"askel\"ainen and Stenholm \cite{Jaaskelainen:03}. They
determined the conditions under which the movement of a matter wave
can be considered adiabatic in a curved waveguide and developed a
formalism based on localised and de-localised basis states. Here we
will take a more straightforward approach and present a numerical
solution to the process, which will show that despite the existence of
velocity-dependent potentials due to the curvature of the waveguides
\cite{Jaaskelainen:03} the STIRAP process can be observed with high
fidelity.

Our simulations start with a well-localised wave-packet far away from
the coupling area. In time, however, this packet will disperse along
the waveguide, making it hard to quantify the success of the transfer
process. To overcome this problem we introduce an additional harmonic
potential of frequency $\omega_l$ along the $y$-axis, which will lead
to a refocussing of the wave-packet in the longitudinal direction after
a time of $\omega_l/2$. The initial state of our wave-packet is given
by the ground state of an isotropic trap of the transverse frequency
of the waveguide and its movement along the guide is induced by the
harmonic potential as well.

\begin{figure}[tb]
\centering
  \includegraphics[width=0.5\linewidth]{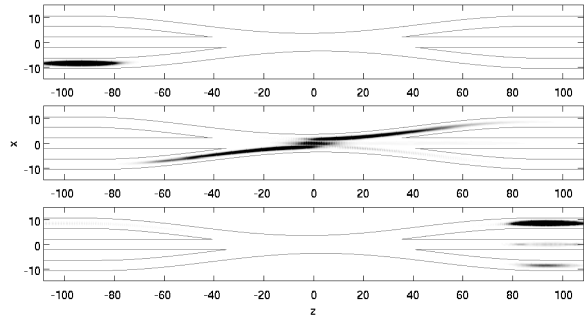}
  \caption{Time evolution of the wave-packet in a counter-intuitive
    arrangement of the waveguides at three different times for values
    of A=20 and B=0.5. The shapes of the waveguides are indicated by
    the black lines and they are fully separated at the energy of the
    wave-packet in the coupling zone (not visible).}
  \label{fig:intuitive}
\end{figure}

In Fig.~\ref{fig:intuitive} we show the evolution of the wavefunction
at different times throughout the process for a counter-intuitive
arrangement of the waveguides. Starting with the wave-packet located in
the lower guide, one can clearly see that a majority of the
probability is transferred into the upper guide. The evolution of the
same initial state in an intuitive arrangement of waveguides (see
Fig.~\ref{fig:intuitive}) shows significantly less transfer.

\begin{figure}[tb]
\centering
  \includegraphics[width=0.5\linewidth]{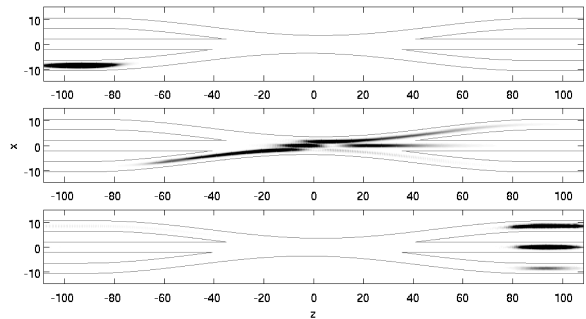}
  \caption{Time evolution of the wave-packet in an intuitive
    arrangements of the waveguides for the same parameters as in
    Fig.~\ref{fig:intuitive}}
  \label{fig:cintuitive}
\end{figure}

\begin{figure}[tb]
\centering
  \includegraphics[width=0.5\linewidth]{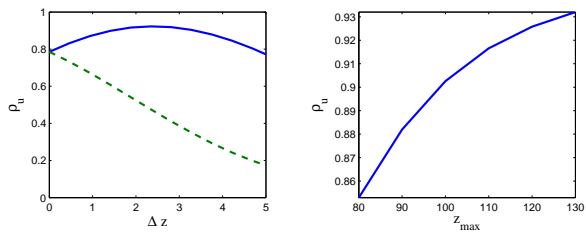}
  \caption{(Left) Probability transferred into the upper waveguide as
    a function of offset between the two outer guides. The full line
    shows the results from the intuitive case and the broken line for
    the counter-intuitive one. (Right) Maximum probability transferred
    as a function of the length of the interaction region.}
  \label{fig:rho}
\end{figure}

The amount of transfer varies as a function of several parameters. The
first one is the distance between the two points of closest approach
of the outer waveguides to the middle one, $\Delta z$. We show the
amount transfered as a function of this quantity in Fig.~\ref{fig:rho}
on the left hand side. The full line (blue) represents the
counter-intuitive case and a maximum at a finite value of $\Delta z$
is visible. The broken line shows the same quantity for the intuitive
setting, clearly indicating that direct tunneling does not lead to
high fidelities.

The second parameter that plays an important role is the degree of
adiabaticity of the process. For a waveguide system this translates
into the velocity with which the atom moves or alternatively the
length of the coupling area. Here we keep the velocity effectively
constant and show on the right hand side of Fig.~\ref{fig:rho} the
variation of the maximum amount transfered as a function of the
overall length of the coupling area. Making the overall structure
longer also corresponds to decreasing the curvature of the waveguides
and thereby reducing the velocity-dependent couplings introduced by it
\cite{Jaaskelainen:03}. As expected we find that a more adiabatic
process leads to a larger transfer probability.

Two caveats have to be pointed out with respect to the above
simulations. While our calculations are carried out with the atom in
the ground state in the transversal direction, this is not a necessary
condition. In fact, the process will work for any state for which
three degenerate asymptotic states exist. This in particular includes
states of higher energy.

Secondly, our simulations are carried out only for the linear case of
a single atom. If one would like to carry out the same process using,
say, a Bose-Einstein condensate, one has to take care of the
non-linearity that arises from the atomic interactions. However, we
believe that our simulations give a very good approximation for low
density condensates or even thermal clouds of atoms.

\section{ Atom  Chips}
\label{sec:AtomChips}

While the above results clearly demonstrate the viability of the
process, it is currently not clear in which experimental system it
will be possible to observe it. One of the problems is that the
asymptotic eigenstates of the system have to be in resonance at any
point in time. This is hard to achieve in many realistic systems as
neighbouring trapping potentials usually strongly influence each other
when they are close enough to allow for significant tunneling rates.

Atom chips are well developed experimental tools these days and
consist of an arrangement of current carrying wires mounted on a
surface \cite{Folman:02}. A current, $I_w$, flowing through a wire
creates a magnetic field around it with the minimum sitting on the
wire.  When applying a homogeneous bias field, $B_b$, in the direction
orthogonal to the wire, a two-dimensional field minimum above the wire
can be created at a height given by \cite{Folman:02}
\begin{eqnarray}
 r_0=\left(\frac{\mu_0}{2\pi}\right)\frac{I_w}{B_b}\;.
\end{eqnarray}
To lift the energetic degeneracy between trapped and untrapped spin
states and thereby avoiding spin flip losses at the field minimum, it
is necessary to apply a second small $B$-field component, $B_{ip}$,
along the axis of the wire (z-axis). This changes the potential at the
minimum from linear to harmonic \cite{Folman:02}
\begin{eqnarray}
  U(r,z)\approx U_z+\frac{1}{2}m\omega^2_r(r-r_0)^2\;,
\end{eqnarray}
where $U_z=m_Fg_F\mu_B|B_{ip}|$ and the radial harmonic trap frequency
is
\begin{eqnarray}
  \omega_r=\frac{\mu_0}{2\pi}\frac{I_w}{r_0^2}\sqrt{\frac{m_Fg_F\mu_B}
    {m B_{ip}}}\;.
\end{eqnarray}
We simulate the STIRAP process by considering three such wires
separated by a distance of 9$\mu m$ initially. The overall length of
the coupling zone chosen such that in the intuitive case several Rabi
oscillations can be expected and the distance between two wires at the
point of closest approach is chosen as 4.5$\mu m$. The applied bias
field has a magnitude of $B_b=100G$ and because of the small curvature
of the wires can be regarded as orthogonal at any point. Since a large
ground state is advantageous for tunnelling the atomic species we
consider is $^{6}$Li.

In general the central minimum will be influenced by the fields from
the two outer wires and increasingly so as the wires come closer. This
will effect the resonance condition and ultimately prevent the STIRAP
process from working. In order to minimize this behaviour we make use
of a trick and adjust the current going through the middle wire to be
slightly lower than the ones going through the outer wires. In our
simulations we choose $I_m=700$mA for the middle wire and
$I_{l,r}=1000$mA for the two outer wires.

\begin{figure}[tb]
\centering
  \includegraphics[width=0.7\linewidth]{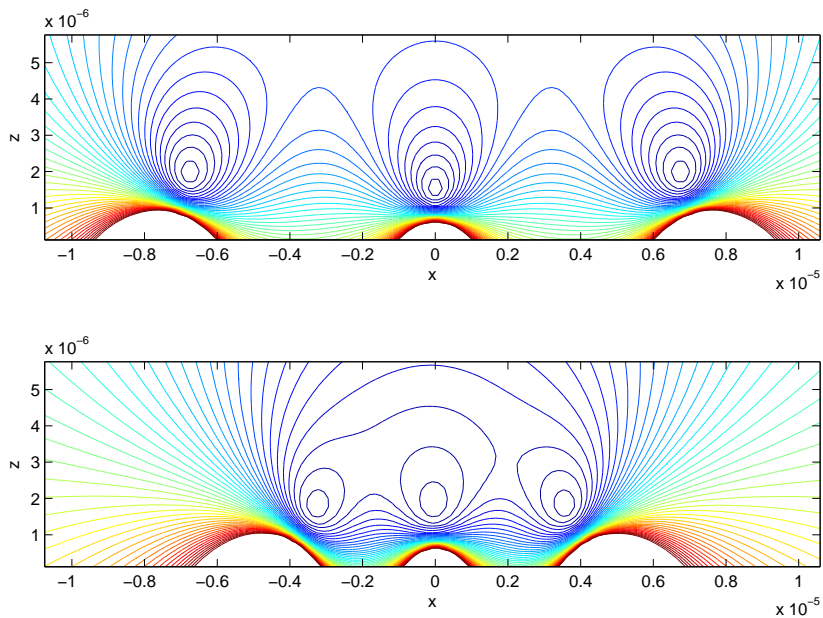}
  \caption{Potentials above the three wires on an atom chip when all
    wires are separated at equal distance (upper) and at the point
    where the wire on the left is closest to the centre wire (lower)}
  \label{fig:AC}
\end{figure}

Fig.~\ref{fig:AC} shows the potential above the wires for the two
different situations of symmetric distance between all wires (upper
graph) and when the left wire is closer to the centre one than the
right wire (lower graph). While an asymmetry in the second case is
clearly visible, its effect on the potentials is moderate.

A full 3D simulation of the STIRAP process in these potentials is a
numerically taxing task and beyond our current capabilities.  We have
therefore simulated the process by using the two-dimensional
potentials of the kind displayed in Fig.~\ref{fig:AC} and changing the
distance between the wires as a function of time. In doing so we
neglect the dispersion of the wavefunction along the longitudinal
direction. However, since we have shown in Section \ref{sec:Model}
that the dispersion does not have any significant effect on the
transfer fidelity, our simulations can be seen as a good approximation
to the full situation.

\begin{figure}[tb]
\centering
\includegraphics[width=0.8\linewidth]{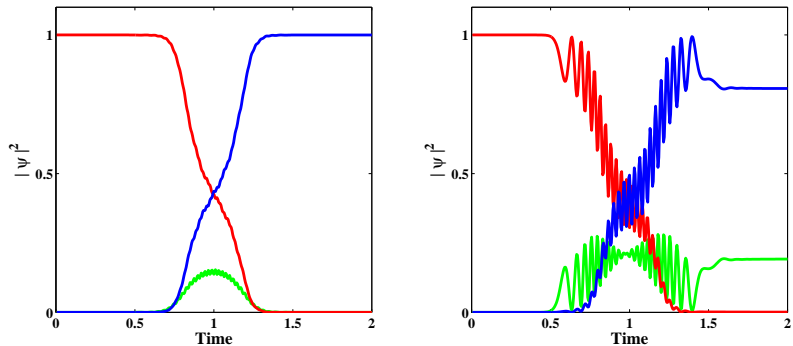}
\caption{Population in the individual waveguides as a function of time
  for the counter-intuitive (left) and intuitive (right) waveguide
  arrangements. The population in the trap on the left is shown by the
  blue line, the middle on by the green line and the one on the right
  by the red line.}
\label{fig:population}
\end{figure}

In Fig.~\ref{fig:population} we show the results of these simulations
by displaying the populations in the individual traps as a function of
time for the intuitive (right) and the counter-intuitive case
(left). Initially all population is on the left hand side and it can
be clearly seen that in the counter-intuitive situation there is a
smooth transition over to the right hand side. While in the perfect
STIRAP setup no population should ever appear in the central trap, the
various imperfections of this realistic example lead to a finite
occupation during the process. However, at the very end no population
is left in the middle trap. Contrary to this, the graph for the
intuitive case shows Rabi oscillations between neighbouring waveguides
and a less than full transfer of the wavefunction. These are two signs
that would allow to distinguish adiabatic transfer from simple
tunneling. 

The fact that we achieve higher transfer fidelities in this
non-perfect situation compared to the results presented in Section
\ref{sec:Model} is purely due to being able to evolve more
adiabatically in time than in space due to the limitations of our
computer hardware.

\section{Conclusion}
We have investigated the use of the STIRAP technique to transfer
atomic wave-packets between neighbouring waveguides. Using an
idealised system, we have first shown that the dispersion along the
guide does not significantly affect the transfer probability. This was
done by introducing a harmonic potential along the longitudinal axis,
which allowed to refocus the wave-packet after half an oscillation
period.  We have then simulated the STIRAP process using realistic
potentials created above current-carrying wires on atom chips and
shown that by chosing a lower current for the central wire the
energetic resonance condition can be fulfilled at any point to a very
high degree. The results clearly showed that adiabatic transfer in the
counter-intuitive setup leads to a higher fidelity and can be clearly
distinguished from direct tunneling in the intuitive setup by the
absence of Rabi oscillations.

\ack This project was supported by Science Foundation Ireland under
project number 05/IN/I852. BOS acknowledges support from IRCSET
through the Embark Initiative RS/2006/172. The work is dedicated to
Prof.~Stig Stenholm on his 70$^{th}$ birthday.

\end{document}